\documentclass{iaus}
\usepackage{graphicx}

\newcommand{\HI}{H\,{\sc i}}
\newcommand{\himass}{{HIZOA\,J0836$-$43}}
\newcommand{\msun}{\,{\rm M}_\odot}
\newcommand{\lsun}{\,{\rm L}_\odot}
\newcommand{\kms}{{\,km\,s$^{-1}$}}
\def\la{\mathrel{\hbox{\rlap{\hbox{\lower4pt\hbox{$\sim$}}}\hbox{$<$}}}}
\def\ga{\mathrel{\hbox{\rlap{\hbox{\lower4pt\hbox{$\sim$}}}\hbox{$>$}}}}

\def\arcsec{\hbox{$^{\prime\prime}$}}

\def\fm{\hbox{$.\!\!^{\rm m}$}}

\def\fdg{\hbox{$.\!\!^\circ$}}

\def\farcs{\hbox{$.\!\!^{\prime\prime}$}}

\def\micron{\hbox{$\mu$m}}
\def\deg{{^\circ}}

\title[~~ The nearby \HI-massive LIRG HIZOA\,J0836$-$43] %% give here short title %%
{The SED of the nearby \HI-massive LIRG HIZOA\,J0836$-$43:\\ from the NIR to the radio domain}

\author[Ren\'ee C. Kraan-Korteweg \& Michelle E. Cluver]   %% give here short author list %%
{Ren\'ee C. Kraan-Korteweg$^1$
 \and Michelle E. Cluver$^2$\thanks{Present address: AAO, Epping NSW 1710, Australia},
}

\affiliation{$^1$Astronomy Department, Astrophysics, Cosmology and Gravity Centre (ACGC), University of Cape Town, Rondebosch 7700, South Africa
\\ email: {\tt kraan@ast.uct.ac.za} \\[\affilskip]
$^2$IPAC, California Institute of Technology, Pasadena, CA 91125, USA
\\email: {\tt mcluver@aao.gov.au}}

\pubyear{2011} %% 
\volume{284}  %% insert here IAU Symposium No.
\pagerange{1--12}
\jname{The Spectral Energy Distribution of Galaxies}
\editors{R.J. Tuffs \&  C.C.Popescu, eds.}
\begin{document}

\maketitle

\begin{abstract}
  \himass\ is one of the most \HI-massive galaxies in the local
  ($z<0.1$) Universe. Not only are such galaxies extremely rare, but
  this ``coelacanth" galaxy exhibits characteristics -- in particular
  its active, inside-out stellar disk-building -- that appear more
  typical of past ($z \sim 1$) star formation, when large gas
  fractions were more common.  Unlike most local giant \HI\ galaxies,
  it is actively star forming. Moreover, the strong infrared emission
  is not induced by a merger event or AGN, as is commonly found in
  other local LIRGs. The galaxy is suggestive of a scaled-up version
  of local spiral galaxies; its extended star formation activity
  likely being fueled by its large gas reservoir and, as such, can aid
  our understanding of star formation in systems expected to dominate
  at higher redshifts. The multi-wavelength imaging and spectroscopic
  observations that have led to these deductions will be
  presented. These include NIR ($JHK$) and MIR (Spitzer;
  $3-24\micron$) imaging and photometry, MIR spectroscopy, ATCA
  \HI-interferometry and Mopra CO line emission observations. But no
  optical data, as the galaxy is heavily obscured due to its location
  in Vela behind the Milky Way.

%%Maybe delete last 2 sentences

\end{abstract}

\firstsection % if your document starts with a section,
\section{Introduction of  \himass}

The galaxy \himass, originally discovered in the Parkes deep Multibeam
\HI-survey of the Zone of Avoidance (ZOA; e.g. Kraan-Korteweg et
al. 2005), is a rapidly rotating disk galaxy ($v_{\rm rot}=630$\kms;
$D_{\rm HI} = 130$kpc) with an \HI-mass of $M_{\rm HI} = 7.5 \cdot
10^{10}\msun$ and a dynamical mass of $M_{\rm tot} = 1.4 \cdot
10^{12}\msun$ (Donley et al. 2006). It has a prominent bulge and
smallish stellar disk ($B/D \sim 0.8$) of evolved stars of $M^* = 4.4
\cdot 10^{10}\msun$ embedded in the five times larger HI-disk, while
its current hearty star forming activity makes it a luminous infrared
galaxy (LIRG; next section).

\himass\ is the most \HI-rich galaxy known. Such galaxies are extremely
rare ($\sim 3 \cdot 10^{-8}$/Mpc$^3$) according to the current best
determined local galaxy HI-mass function (HIMF; Zwaan et al. 2005) and
should not have been found at all in the explored volume.
%and only forming now ($z \la 1$) according to hierarchical structure
%formation (Mo et al. 1998). 
Interestingly, the recent, larger volume ALFALFA survey 40\%-data
release (Haynes et al. 2011) identify several, similarly extreme
\HI-massive galaxies in excess of the predicted HIMF number density.
%Contrary to other known giant HI galaxies, 
%like Malin~1 (Bothun et al. 1987; Pickering et al. 1997) 
%this disk galaxy is neither quiescent nor of low surface brightness.
Independently, galaxies with such vast reservoirs of gas were more
common in the past, as were LIRGs. Hence this enigmatic, relatively
nearby ($v=10\,689$\kms) galaxy is an ideal {\sl local} probe that enables
detailed studies of evolutionary processes and the transformation of
gas into stars.

\section{The Observational Data}

The galaxy lies deep in the ZOA ($\ell=262\fdg48, b=-1\fdg64$) and
suffers from severe Galactic foreground extinction ($A_V = 7\fm3$),
rendering it practically invisible in the optical. Hence our detailed
follow-up observations focus on the infrared, radio and mm domain.

{\underline{\it IRSF NIR imaging survey}}. The deep $J H K$-imaging
survey of $2.2\Box\deg$ with the Infrared Survey Facility (IRSF)
%at the 1.4 m Japanese telescope 
in Sutherland (SAAO) aimed at learning more about an environment that
permits such a gas giant to evolve. The survey uncovered 404 galaxies
to the completeness limit of $K_s < 15\fm8$ (Cluver 2009). It finds the
volume around \himass\ to be overdense in sub-$L{^*}$ systems, and none in the $L{^*}$ range.  Hence it seems to live in a tranquil, underdense environment on the edge of a void. There are no
indications of a major merger that could have triggered a starburst
(see also Fig.~2). This may have allowed the formation of its giant
\HI-disk through minor merger or accretion -- and aided its survival.

\begin{figure}[h]
\vspace*{-0.4cm}
\begin{center}
 \includegraphics[width=0.75\textwidth]{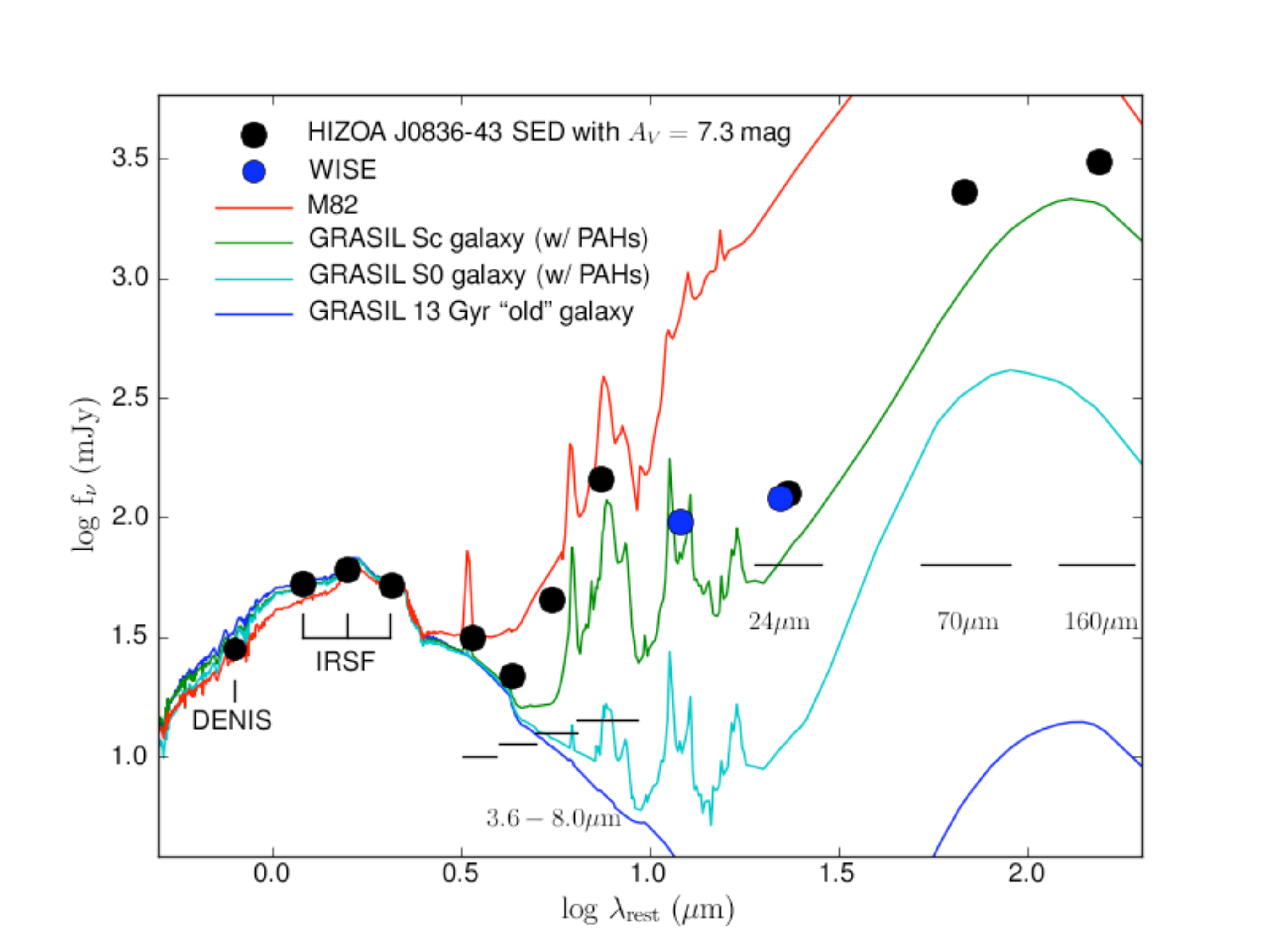} 
\vspace*{-0.3 cm}
\caption{SED based on DENIS $I$, IRSF $J,H,K_s$, Spitzer IRAC
  ($3.6-8.0$\micron) and MIPS (24, 70 \& 160\micron), as well as the
  WISE 12 \& 22\micron\ bands -- all corrected for extinction ($A_V =
  7\fm3$). Superimposed are GRASIL model templates and the M82
  spectrum for comparison. While the high MIR/FIR luminosity makes
  this galaxy a LIRG, it clearly is not consistent with a
  starburst. (Adapted from Cluver et al. 2008).}
   \label{SED}
\end{center}
\end{figure}

{\underline{\it Spitzer IRAC and MIPS imaging and spectroscopy}}.  The
photometry from the Spitzer IRAC and MIPS band was presented in Cluver
et al. 2008. It is reproduced in the SED (Fig.~\ref{SED}) together
with the IRSF data, the DENIS $I$-band
(Donley et al. 2006) and, now available, WISE 12 \& 22\micron\ fluxes (Jarrett,
priv. comm.).

The resulting total infrared luminosity $L_{\rm TIR} = 1.2 \cdot
10^{11}\lsun$ defines it as a LIRG. Despite
the high star formation rate (SFR$ = 20.5\msun/$yr),
%; derived according the Kennicutt 1998), 
the comparison of its SED to various galaxy model templates, and the
archetypal local starburst M82 %(Sturm et al. 2000)
in Fig.~1, shows that \himass\ is neither a starburst galaxy nor -- as
originally assumed from its NIR morphology -- a S0/Sa-galaxy. It
rather resembles a Sc galaxy with strong emission in the mid-IR
$5-20\micron$ range, due to the prevalence of PAH molecules, and a
prominent FIR cold dust continuum. The latter is substantiated by the
mid-IR spectroscopy (Cluver et al. 2010) which shows particularly
luminous 6.2 and 7.7\micron\ PAH features (stronger than typical
starbursts; e.g. Brandl et al. 2006) in contrast to the continuum
emission from warm dust. The latter is surprisingly weak (i.e. lower
than any of the local LIRGs in GOALS), suggestive of an extended dust
geometry. Cold dust is plentiful, however, and a dominant contributor
to the total infrared luminosity with its steeply rising FIR emission
for $\lambda \ga 60\micron$.

The IRS spectroscopy of the nucleus further reveals strong excited
nebular lines (e.g. [Ne\,{\sc ii}], [S\,{\sc iii}], [Si\,{\sc ii}],
[Ar\,{\sc ii}]). While these are typical of starburst regions, the
line ratios indicate a softer radiation field (like PDR's) and AGN
activity is weak or absent. This is independently confirmed by the
absence of [Ne\,{\sc v}] and [O\,{\sc iv}] lines.
%Also apparent in the spectrumare weak rotational H$_2$ lines whose line ratios suggest a temperature of $\sim 330$K and a warm molecular gas mass of only $M_{\rm H_2} = 1.3 \cdot 10^7\msun$. While this not is not reallyunusual for LIRGs, it is very low compared to the amount of cold neutral gas (0.02\%).

{\underline{\it ATCA and MOPRA observations}}.  With the surprising
dearth of warm dust despite the strong star
forming activity (PAH's) and the large reservoir of cold \HI\ gas, Mopra
CO line observations (1-0) were used to probe the cold molecular gas content. The spectrum shows a double horn signal, matching the \HI-profile width perfectly. %without a signature around the systemic velocity though. 
However, the total molecular gas mass
(H$_2 + $He) of $3.9 \cdot 10^9\msun$ is lower by a factor of three
than expected from the infrared luminosity, and only a meager 5\% compared to the atomic \HI\ (Cluver et al. 2010). 
%Though  the beam did not cover the whole galaxy.
Both the total and cold molecular gas fraction (gas/gas+stellar mass)
are high (64\% and 8\%) compared to the mean of local galaxies (e.g.
20\% total; Leroy et al. 2008, and 6\% molecular for the CO-detected
stellar-selected massive galaxies; Saintonge et al. 2011), but low
compared to the local LIRG sample (Wang et al. 2006). Such high gas
fraction values are the standard though at higher redshifts
(e.g. Daddi et al. 2010 for their NIR selected $BzK$ sample at $z\sim
1.5$).

\begin{figure}[b]
% \vspace*{-2.0 cm}
\begin{center}
 \includegraphics[height=4.5cm]{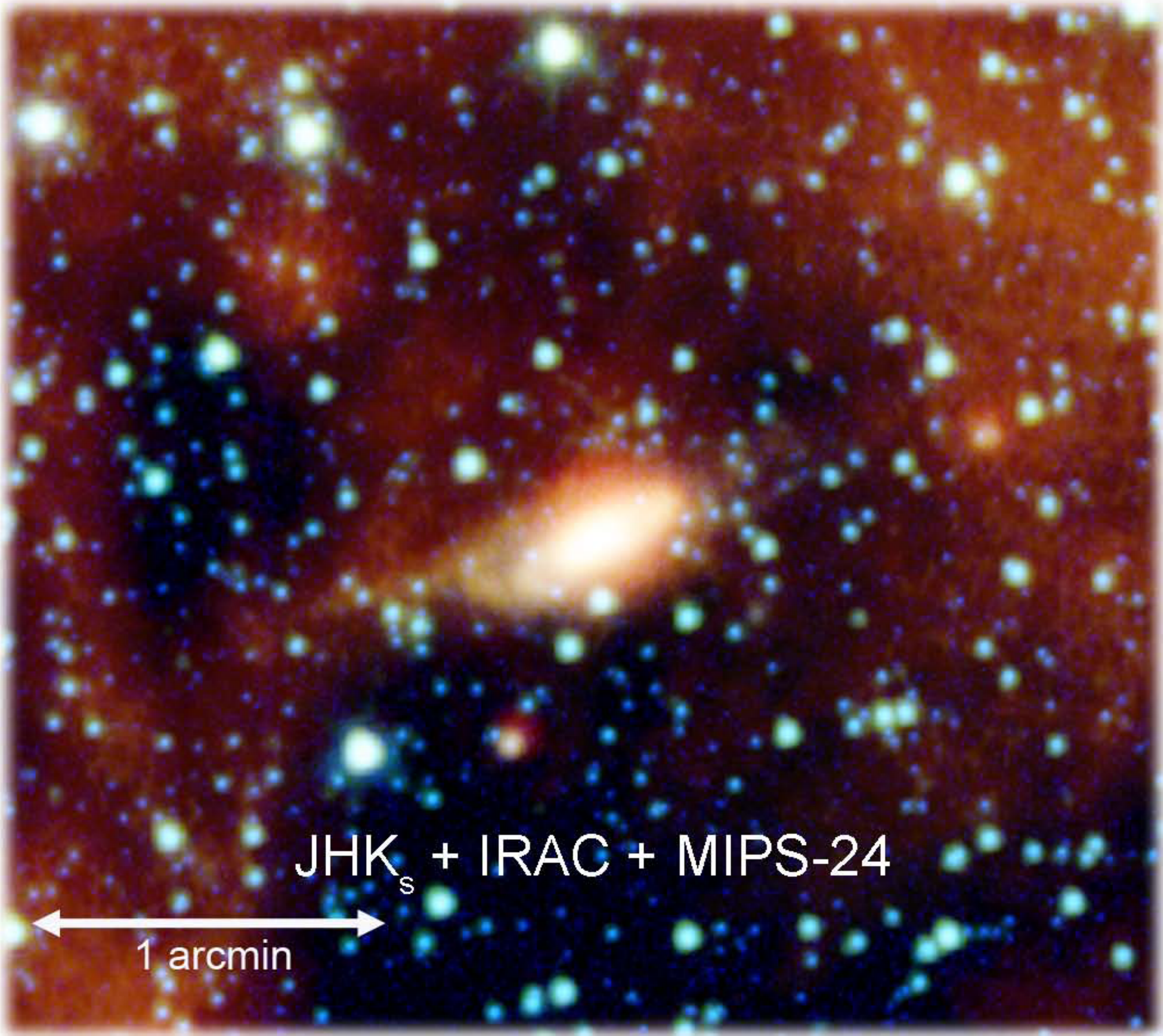} 
\includegraphics[height=4.5cm]{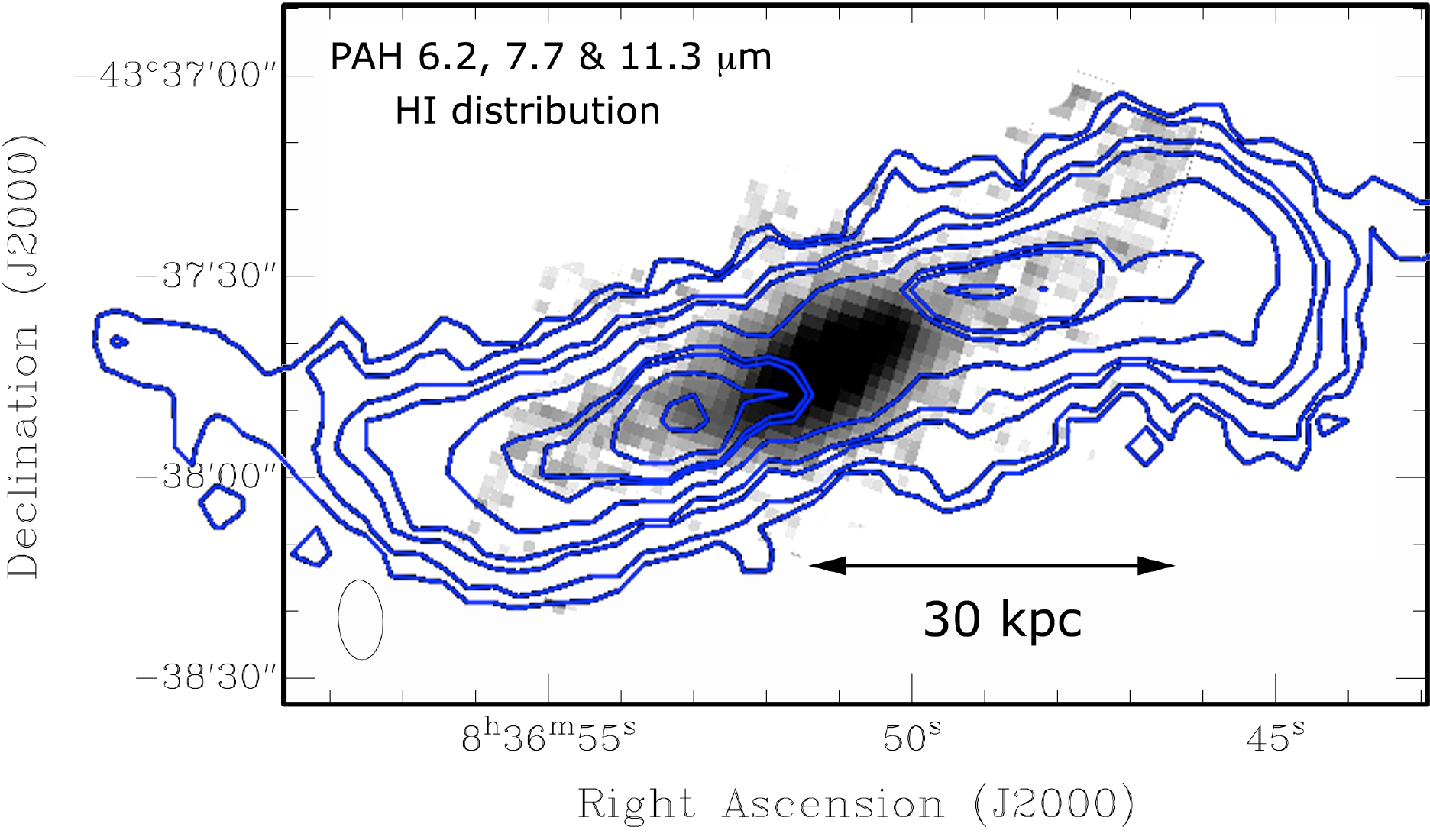}
% \vspace*{-1.0 cm}
\caption{Left Panel: Composite of $J,H,K_s$ and Spitzer IRAC \&
  MIPS 24\micron\ images. Right panel: \HI\ distribution superimposed
  on composite spectral maps of the 6.2, 7.7, and 11.3\micron\ PAH
  emission bands from Spitzer IRS. Note the asymmetry of PAH
  emission on the left (beyond the edge of the stellar disk),
  coinciding with an enhancement seen in the \HI.}
   \label{fig2}
\end{center}
\end{figure}
Given this dichotomy, it seemed of importance to learn more about the
interplay between star formation activity and gas physics in the
galaxy itself. This required longer baseline observations with ATCA to
complement the earlier Donley et al. (2006) data, which only had
30\arcsec\ angular resolution. Of interest is the connection -- if any
- with the star forming region, which extends beyond the old stellar
bulge, suggestive of inside-out stellar disk-building, and exhibits a
warp towards the eastern side. This is evident in the composite NIR \&
MIR image (left panel of Fig.~2) as well as the composite spectral map
($2\arcsec$ resolution) of the PAH 6.2, 7.7 and 11.3\micron\ features
(right panel). Note that the latter was obtained from SL
(5--14\micron) IRS observations mapping across the IR disk.

A preliminary \HI-map from the newly acquired data is superimposed on
the right panel in Fig.~2 for comparison. We note that the improved
spatial resolution reveals an enhancement in the atomic gas exactly at
the location of the warp. This may be the result of a recent tidal
interaction which has triggered star formation as gas flows towards
the center of the galaxy. This \HI-data will be explored in detail in
Cluver et al. (in prep).

\section{Connection to evolutionary processes at higher redshifts}

How does this galaxy relate to systems at varying redshift/evolutionary stages? Albeit at the extreme end, \himass\ seems to fall exactly on the \HI-mass and SFR relation defined by the nearby SINGS galaxy sample (contrary to M82 or Malin~1). This suggests it to be a scaled-up version of local disk galaxies. Comparing it to the local LIRG sample ($z \la 0.1$; Wang et al. 2006) its specific starformation rate of 0.47\,Gyr$^{-1}$ appears to follow the overall trend. But, it again lies at an extreme stage, implying active stellar building; in an additional 2 Gyrs of time it will double its stellar mass and fit the `current' local LIRG data. This hint of relic SF properties is also seen when comparing \himass\ to the higher redshift SF galaxy sample by e.g. Bell et al (2005), where it fits well on sSFR-relation of $z \sim 0.7$ galaxies, but does not conform at all to the local SF galaxy sample. Idem ditto when compared to the Genzel et al. (2010) sample where it closely aligns with the $z\sim 1.2$ sample. Tellingly, when entering the parameters of \himass\ in the so-called main sequence of SFR as a function of stellar mass and redshift (Bouch\'e et al 2010) a redshift (age) of $z \sim 0.95$ is returned, suggesting a tantalising link to a distant epoch of star formation.

\himass\ displays a combination of high gas fraction with low efficiency, but extended, star formation in a low excitation disk, also found in the sample of normal, near-IR selected ($BzK$) galaxies at $z\sim1.5$ of Daddi et al. (2010), who consider them  scaled-up local disk galaxies.
Since larger gas fractions at higher redshift permit larger $L_{\rm IR}$ before invoking special events like mergers that show an offset in correlations with $L_{\rm IR}$ (Nordon et al. 2011), \himass\ can be studied as an analogue of disk building processes dominating at the peak of stellar mass growth. 

The lower fraction of molecular gas in \himass\ compared to detected high-redshift galaxies (Tacconi et al. 2010) highlights the need for an accurate molecular gas inventory, in parallel to increased high $z$ detections. Though, as pointed out by Groves at this conference, ``dark" molecular gas may be a feature of the active ISM enrichment occurring in the disk. %and the resulting metallicity gradient.
Herschel and ALMA is needed to provide a cohesive picture that can be used to understand disk building in gas-rich systems.

\smallskip
\noindent {\bf Acknowledgment} -- RKK thanks the South African National Research Foundation and the University of Cape Town for financial support.

\newpage

\noindent {\bf Sukanya Chakrabarti}:
What is your interpretation of the high IR emission? My interpretation is that this is due to a minor merger. We now have scaling relations that will allow you to derive the satellite mass from the \HI-map. So I suggest that you calculate the Fourier amplitude.

\medskip

\noindent {\bf Response}:
Thank you. That is an interesting idea for the future, but we will
await the reduction of our new \HI-data and its analysis first. The
current \HI-data maps are quite smooth over most of its gas disk
without any kinks in its contours. This makes the minor-merger scenario
unlikely; at least, it could not have occurred over the last 2 Gyr or
so to account for such a stable disk. Secondly, while a minor merger
(or instability caused by the accretion of a satellite) might explain
the strong IR luminosity due to triggered star formation/nuclear
starburst, it does not account for the extreme mass in \HI-gas in
\himass, nor the abnormally high gas fraction. In addition, the latter
scales beautifully with normal -- albeit higher redshift --
starforming galaxies. What remains unresolved as yet is the question,
how \himass\ could have acquired/accreted all this cold gas.

\bigskip

\noindent {\bf Martin Bureau}:
The comparison of the CO and \HI in the spectra shows a lack of CO (thus H$_2$) in the rising part of the rotation curve, hence central part of the galaxy. How can this be reconciled with signatures of star formation in the centre, e.g. in the nuclear infrared spectrum shown.

\medskip

\noindent {\bf Response}:
Note that the "nuclear spectrum" covers the "nuclear region" and is a $9\farcs25$ extraction, so it picks up more than just the nucleus itself. Secondly, the central low-velocity gas might well be present at a lower level compared to the \HI-gas distribution (i.e. not detectable with the current sensitivity of our Mopra observations). This is seen quite often in other spiral galaxies when comparing \HI\ and CO observations. Much of the very central CO may also have been used in a previous epoch of star formation which produced the current old stellar bulge (which has a disk scale length of 4~kpc, respectively 7\arcsec). But interesting is that the steep rotation in the \HI\ velocity appears matched to the CO, so the central region achieves a high velocity close in. This may be important for extended building into the disk. We really need spatial information accompanied by velocity resolution (as we now have with \HI) to understand this, and ALMA will make this possible. Also, given the prominence of [Si\,{\sc ii}] in the spectrum, it is not inconceivable that some of the CO is being photodissociated, whereas the molecular hydrogen is shielded, making it ``dark''. Herschel (proposal submitted) in combination with ALMA (planned) will enable a detailed study of such a mechanism.

\end{document}